# Electronic phase separation in iron selenide (Li, Fe)OHFeSe superconductor system

Yiyuan Mao(毛义元)[1,2,3]†, Jun Li(李军)[4]†, Yulong Huang(黄裕龙)[1,2,3]†, Jie Yuan(袁洁)[1,3], Zian Li(李子安)[1,3], Ke Chai(柴可)[5], Mingwei Ma(马明伟)[1,3], Shunli Ni(倪顺利)[1,2], Jinpeng Tian(田金鹏)[1,2], Shaobo Liu(刘少博)[1,2], Huaxue Zhou(周花雪)[1], Fang Zhou(周放)[1,2], Jianqi Li(李建奇)[1,2,3], Guangming Zhang(张广铭)[6], Kui Jin(金魁)[1,2,3], Xiaoli Dong(董晓莉)[1,2,3]*, and Zhongxian Zhao(赵忠贤)[1,2,3]*

[1]Beijing National Laboratory for Condensed Matter Physics and Institute of Physics, Chinese Academy of Sciences, Beijing 100190, China

[2]University of Chinese Academy of Sciences, Beijing 100049, China

[3]Key Laboratory for Vacuum Physics, University of Chinese Academy of Sciences, Beijing 100049, China

[4]Research Institute of Superconductor Electronics, Nanjing University, Nanjing 210093, China

[5]School of Physics, Beijing Institute of Technology, Beijing 100081, China

[6]State Key Laboratory of Low Dimensional Quantum Physics and Department of Physics, Tsinghua University, Beijing 100084, China

† These authors contributed equally to this work

* E-mail: dong@iphy.ac.cn (X. L. D.); zhxzhao@iphy.ac.cn (Z. X. Z.)

**Abstract** The phenomenon of phase separation into antiferromagnetic (AFM) and superconducting (SC) or normal-state regions has great implication for the origin of high-temperature (high-$T_c$) superconductivity. However, the occurrence of an intrinsic antiferromagnetism above the $T_c$ of (Li, Fe)OHFeSe superconductor is questioned. Here we report a systematic study on a series of (Li, Fe)OHFeSe single crystal samples with $T_c$ up to ~41 K. We observe an evident drop in the static magnetization at $T_{afm}$ ~125 K, in some of the SC ($T_c <$ ~38 K, cell parameter $c <$ ~9.27 Å) and non-SC samples. We verify that this AFM signal is intrinsic to (Li, Fe)OHFeSe. Thus, our observations indicate mesoscopic-to-macroscopic coexistence of an AFM state with the normal (below $T_{afm}$) or SC (below $T_c$) state in (Li, Fe)OHFeSe. We explain such coexistence by electronic phase separation, similar to that in high-$T_c$ cuprates and iron arsenides. However, such an AFM signal can be absent in some other samples of (Li, Fe)OHFeSe, particularly it is never observed in the SC samples of $T_c >$ ~38 K, owing to a spatial scale of the phase separation too small for the macroscopic magnetic probe. For this case, we propose a microscopic electronic phase separation. It is suggested that the microscopic static phase separation reaches vanishing point in high-$T_c$ (Li, Fe)OHFeSe, by the occurrence of two-dimensional AFM spin fluctuations below nearly the same temperature as $T_{afm}$ reported previously for a (Li, Fe)OHFeSe ($T_c$ ~42 K) single crystal. A complete phase diagram is thus established. Our study provides key information of the underlying physics for high-$T_c$ superconductivity.

High-$T_c$ superconductivity in cuprates, derived from an antiferromagnetic (AFM) Mott insulator through carrier doping, tends to coexist with spin or charge density wave orders in microscopic to macroscopic scales from dilute carrier doping. Such electronic phase separations, in which hints for the superconducting (SC) pairing are embedded, have attracted extensive attention theoretically and experimentally over the past decades.[1-8] Similar electronic phase separation in iron arsenide superconductors is also significant.[9-15] However, it is still far from clear the roles of spin or orbital degree of freedom in the high-$T_c$ superconductivity of multiband iron-based materials. In iron selenide FeSe-122 superconductors like $K_y Fe_{2-x} Se_2$ ($T_c$ ~30 K), in particular, the situation becomes more complicated. Distinct phases are present in $K_y Fe_{2-x} Se_2$ by high-resolution transmission electronic microscopy,[16] and the SC phase is always inter-grown with an *extrinsic* AFM insulating $K_2 Fe_4 Se_5$ (245) phase. Such unavoidable chemical and structural phase separations[16-18] hamper the study on the intrinsic electronic property of material. In the simplest binary FeSe superconductors ($T_c$ ~9 K), on the other hand, there appears a structural transition at ~90 K and no long-range magnetic order occurs in the bulk material, though in the parent monolayer film of FeSe an AFM order was observed below ~140 K.[19]

In contrast to the prototypal FeSe and FeSe-122 superconductors, the recently discovered iron selenide intercalate of (Li, Fe)OHFeSe (FeSe-11111)[20] is free from the complications of the chemical phase separation (without the 245 phase)[21] and structural transition. Moreover, it shows a high $T_c$ over 40 K under ambient conditions, even above 50 K under a 12.5 GPa pressure,[22] and a Fermi topology similar to the high-$T_c$ (>65 K) FeSe monolayer.[23,24] Importantly, in a recent study we have observed an appreciable decrease in the magnetization at ~125 K in non-superconducting (non-SC) (Li, Fe)OHFeSe powder.[21] Furthermore, by a subsequent study on an optimal ($T_c$ ~42 K) (Li, Fe)OHFeSe single crystal,[25] we have shown that the normal-state electronic behaviors in the FeSe-layers of (Li, Fe)OHFeSe are getting highly two-dimensional (2D) and AFM spin fluctuations (AFM-SF) set in, below nearly the same temperature (~120 K) as that of the magnetic drop mentioned above. Most recently, density functional calculation[26,27] also suggests the presence of AFM order within the superconducting FeSe-layers of (Li, Fe)OHFeSe. Thus, (Li, Fe)OHFeSe turns out to be an ideal system for investigating the intrinsic electronic phase separation and the interplay of magnetism and high-$T_c$ superconductivity in iron-based family. However, a recent neutron diffraction on a non-SC deuterated (Li, Fe)ODFeSe[28] sample (without the ~125 K AFM signal ) did not detect any long-range magnetic order, though the observation of spin resonance was reported.[29,30] It was thus speculated[28] that the AFM signal at ~125 K reported for the hydroxide (Li, Fe)OHFeSe powder might be caused by a so-called Verwey transition at ~120 K of $Fe_3O_4$, plausibly present as an impurity.

In this letter, we report a systematic study on a series of high-quality superconducting (~20 K < $T_c$ < ~41 K) and non-superconducting (Li, Fe)OHFeSe single crystal samples. In some of the SC ($T_c$ < ~38 K) and non-SC samples, we observe an evident drop in the magnetization at an almost constant temperature scale ($T_{afm}$) of ~125 K. In addition, a corresponding upward kink at $T_{afm}$ is visible in the in-plane electrical resistivity for some of the (Li, Fe)OHFeSe samples. It is shown that this AFM signal is intrinsic to (Li, Fe)OHFeSe and no impurity phases like $Fe_3O_4$ appear. The magnetic and electrical transport measurements give macroscopic properties of the material. Hence, our experiments indicate the coexistence, in a mesoscopic to macroscopic scale, of an AFM state with the normal (below $T_{afm}$) or superconducting (below $T_c$) state in (Li, Fe)OHFeSe. Such coexistence can be explained by electronic phase separation. The nearly constant AFM transition temperature ($T_{afm}$) is a common feature to electronic phase separations. However, the magnetic signal of the AFM phase below $T_{afm}$ can be imperceptible on some other samples of $T_c < $ ~38 K, particularly it is never observed on the SC samples of $T_c > $ ~38 K. It is because of varying scales of the phase separation from sample to sample, suggested by a positive correlation between the SC Meissner and AFM signal sizes. For these samples showing no such an AFM transition, we propose a microscopic picture of the electronic phase separation. This microscopic static-phase separation reaches vanishing point in high-$T_c$ (~42 K) (Li, Fe)OHFeSe, suggested by the previously reported 2D AFM spin fluctuations occurring below nearly the same temperature as $T_{afm}$ in the FeSe-layers.[25] Thus, we establish a complete electronic phase diagram for (Li, Fe)OHFeSe superconductor system.

The (Li, Fe)OHFeSe single crystals were synthesized by the hydrothermal ion-exchange technique that we developed and first reported elsewhere.[25] The X-ray diffraction (XRD) measurements were performed at room temperature on an 18 kW MXP18A-HF diffractometer with Cu-K$_a$ radiation, using a $2\theta$ range from 5° to 80° and a $2\theta$ scanning step of 0.01° (single crystal) or 0.02° (powder). The in-plane electrical resistivity is measured on a Quantum Design PPMS-9. The dc magnetic measurements were carried out on a SQUID magnetometer (Quantum Design MPMS XL-1). The electron energy-loss spectroscopy (EELS) technique combined with a transmission electron microscope was used for local probing of the composition and spectroscopic information of the specimens. The transmission electron microscope (TEM, ARM200F, JEOL Ltd.) was equipped with a spherical aberration corrector (CEOS GmbH).

Fig. 1(a) shows the XRD patterns for all the SC and non-SC (Li, Fe)OHFeSe single crystal samples. Three types of the sample names are used for the superconducting single crystals, corresponding respectively to their distinct AFM signal sizes (detailed characterizations given below). (1) In the samples denoted by the names of S41, S40, S38 and S28, the AFM signal is imperceptible by magnetic measurement; (2) in the samples SA37, SA26 and SA20, the AFM signal is evident; and (3) the samples SA'38 and SA'24 exhibit an AFM signal size intermediate between the S- and SA-samples of similar $T_c$. The numbers in the sample names stand for the $T_c$ values. The sample NSC is non-superconducting. All the SC and non-SC samples display a single preferred crystal orientation of (001). From the zoom-in (006) Bragg reflections shown in Fig. 1(b), a left shift of the peak position with increasing $T_c$ is clearly visible. This indicates a positive correlation between the $T_c$ and the interlayer separation of (Li, Fe)OHFeSe, consistent with our previous reports.[21,31] The powder XRD patterns are given in Fig. 1(c) for some of the (Li, Fe)OHFeSe single crystals. All the reflections in each powder XRD pattern can be well indexed on the known tetragonal structure (space group *P4/nmm*) for (Li, Fe)OHFeSe. No impurity phases like $Fe_3O_4$ can be detected by the powder XRD. The calculated lattice parameters of *a* and *c* (Table 1) are in agreement with our earlier results.[21,25]

Fig. 2(a) shows three characteristic oxygen K edges (electronic excitation from 1s to 2p in oxygen ions) in the electron energy-loss spectra for the respective (Li, Fe)OHFeSe (SA37), $Fe_3O_4$, and FeOOH samples. The spectral features for the sample SA37, showing an evident AFM transition at $T_{afm}$, bear no resemblance to those of $Fe_3O_4$ and FeOOH. It clearly indicates the absence of $Fe_3O_4$ and FeOOH impurities. We note that such EELS measurements were performed using a finely-focused electron beam with a size of ~50 nm for a number of randomly selected crystalline grains. The absence of $Fe_3O_4$ grains in the SA-sample is also confirmed by direct lattice imaging and electron diffraction measurements. Therefore, we further verify that the antiferromagnetism below $T_{afm}$ is intrinsic to (Li, Fe)OHFeSe.

The superconductivity of the single crystal samples are characterized by magnetic susceptibility and confirmed by in-plane electrical resistivity measurements. The data for some representative samples are given in Figs. 3(a) and 3(b) (magnetic susceptibility) and Fig. 2(b) (electrical resistivity). The resulting superconducting transition temperatures are consistent with our previous reports for the powder,[21] single crystal[25] and film[31,32] samples of (Li, Fe)OHFeSe. All the samples exhibit a 100% superconducting shielding. Interestingly, we observe an evident decrease in the static magnetization at a nearly constant temperature scale ($T_{afm}$) of ~125 K, in the superconducting SA-/SA'-samples ($T_c$ < ~38 K) and non-superconducting NSC-sample (Fig. 3(c)). Correspondingly, an upward kink at $T_{afm}$ is visible in the in-plane electrical resistivity curves, for non-SC (inset of Fig. 2(b)) and lower $T_c$ SC (not shown) samples displaying an evident AFM signal of this kind. Both the magnetic and transport measurements probe macroscopic properties of the material. The evident drop in magnetization signifies a three-dimensional (3D) AFM correlation below $T_{afm}$, which causes additional charge scatterings leading to the corresponding upward kink at $T_{afm}$ in the in-plane resistivity. Therefore, our experimental results provide clear evidence for the coexistence of an AFM state with the normal or superconducting state in (Li, Fe)OHFeSe, in a mesoscopic to macroscopic scale. Similar electronic phase separation was extensively studied on high-$T_c$ cuprates,[2,3,6,33,34] and considered as an intrinsic property in iron arsenide superconductors.[12] By contrast, this drop in magnetization is never discernible on the superconducting samples of $T_c$ > ~38 K, and it can be absent in the superconducting samples (e.g., sample S28 in Fig. 3(c)) of $T_c$ < 38 K as well, as will be explained below.

In Fig. 3(d), we plot the AFM signal size, i.e. $\Delta M/H$ near $T_{afm}$, versus the SC Meissner signal size and present the SC transition width by the color spectrum, for the two sets of SA-/SA'-/S-samples with the respective $T_c$ of ~38 K and ~26 K (Figs. 3(a) and 3(b)). We find that, distinct from the S-samples without the AFM transition at $T_{afm}$, the SA-samples showing the appreciable AFM signal sizes exhibit correspondingly much stronger Meissner signals and sharper SC transitions. Accordingly, for the SA'-samples with weaker AFM signals, their Meissner signal sizes and SC transition widths are just intermediate between the SA- and S-samples of the similar $T_c$ (~38 K or ~26 K). This positive correlation between the SC Meissner and AFM signal strengths provides us important hints for varying scales of the phase separation, among the superconducting samples of the same set. The 100% superconducting shielding indicates that the SC phase is connected in the real space. In the SA-samples, both the phase-separated AFM and normal or SC regions should be large enough in scale (a mesoscopic-to-macroscopic length scale, particularly with the scale of SC regions much bigger than the penetration depth) for the magnetic measurement. As a result, their Meissner and AFM signals are strong. We explain accordingly that, in the S-samples, the AFM and normal/SC states survive in microscopic or nanoscopic clusters, with the size of SC clusters comparable to or less than the penetration depth in particular, so that they are microscopically and homogeneously mixed with each other. That can account for our observation for the S-samples of the quite weak Meissner signals and broad SC transitions[35] as well as the imperceptible AFM signal by the macroscopic magnetic probe. Similarly, for the SA'-samples with the Meissner/AFM signal sizes and the SC transition widths intermediate between the SA- and S-samples, the spread of their AFM and normal/SC clusters may be likewise intermediate in scale. Thus, we propose a microscopic picture for the electronic phase separations in the samples showing no AFM transition at $T_{afm}$. Previous work on both the hydroxide (Li, Fe)OHFeSe[21,36] and deuterated (Li, Fe)ODFeSe[28] also show that the presence or absence of the AFM transition at $T_{afm}$ is sample and synthesis condition dependent.

By a recent study on a high-$T_c$ (~42 K) (Li, Fe)OHFeSe single crystal,[25] we have shown that the static magnetic susceptibility at high temperatures obeys a modified Curie-Weiss law, $\chi_m = \chi_0 + \chi_{CW}$. A small value of $\theta$ in the Curie-Weiss term $\chi_{CW} = C/(T - \theta)$ accounts for a magnetic order (or spin glassy behavior) eventually occurring at a much lower temperature (8.5 - 12 K) in the (Li, Fe)OH interlayers[20,37-39]. Intriguingly, on the other hand, its magnetic susceptibility displays an evident downward deviation, but not a drop seen in the SA-/SA'-samples, from the Curie-Weiss behavior, below a characteristic temperature scale (~120 K) nearly the same as $T_{afm}$. Such a deviation corresponds to the two-dimensional AFM spin fluctuations occurring below this characteristic temperature (denoted by $T_{sf}$ here) of ~120 K in the FeSe-layers of (Li, Fe)OHFeSe. Therefore, it is suggested that the above-proposed microscopic phase-separated static AFM state is reaching vanishing point in high-$T_c$ (Li, Fe)OHFeSe.

Finally, we plot the phase diagram for (Li, Fe)OHFeSe in Fig. 4, by the data of $T_c$, $T_{afm}$, and $T_{sf}$ versus the lattice parameter $c$ (Table 1). In the left azure shaded area, the occurrence of a mesoscopic-to-macroscopic (showing a 3D AFM transition at $T_{afm}$) phase separation is sample and synthesis condition dependent. In the case of no such AFM transition at $T_{afm}$, the phase separation is in a microscopic scale. The dashed blue line is an extrapolation from the right 2D AFM-SF (below $T_{sf}$)[25] based on the present observations, overlapping with the microscopic phase separation region.

In conclusion, our experimental observations indicate the mesoscopic or macroscopic coexistence of an AFM state with the normal (below $T_{afm}$) or superconducting (below $T_c$) state in (Li, Fe)OHFeSe, for $T_c < \sim 38$ K and $c < \sim 9.27$ Å. The AFM transition temperature scale ($T_{afm} \sim 125$ K) is almost constant among the samples, a common feature to electronic phase separations. For (Li, Fe)OHFeSe samples showing no AFM transition at $T_{afm}$ by magnetic measurement, we propose a microscopic picture for the electronic phase separation. The occurrence of two-dimensional AFM spin fluctuations below the characteristic temperature $T_{sf}$ ($\sim 120$ K), almost the same as $T_{afm}$, in the FeSe layers suggests that the microscopic static-phase separation nearly vanishes in high-$T_c$ ($\sim 42$ K) (Li, Fe)OHFeSe. Thus, we obtain a complete electronic phase diagram for this iron selenide superconductor system of (Li, Fe)OHFeSe, providing important information of the underlying physics for high-$T_c$ superconductivity. However, the characteristic length scales for the microscopic-to-macroscopic electronic phase separations in (Li, Fe)OHFeSe need further study using other microscopic and ultrafast techniques.

X.L.D would like to thank Dr. Li Yu and Prof. Shiliang Li (IOP, CAS) for helpful discussions. This work was supported by the National Key Research and Development Program of China (Grant Nos. 2017YFA0303003, 2016YFA0300300 and 2015CB921000), the National Natural Science Foundation of China (Grant Nos. 11574370, 11474338, 11674374 and 61501220), the Strategic Priority Research Program and Key Research Program of Frontier Sciences of the Chinese Academy of Sciences (Grant Nos. QYZDY-SSW-SLH001, QYZDY-SSW-SLH008 and XDB07020100), and the Beijing Municipal Science and Technology Project (Grant No. Z161100002116011).

Figure 1

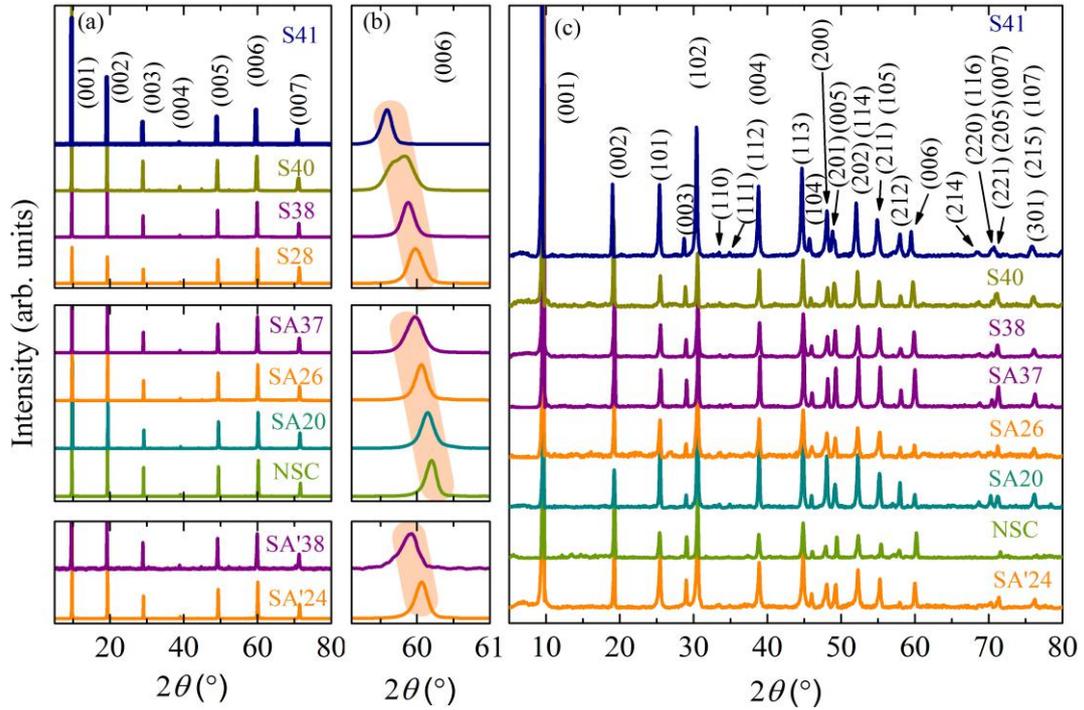

Fig. 1. (a) XRD spectrums for single crystal flakes of (Li, Fe)OHFeSe, all showing a single preferred crystal orientation of (001). (b) Enlarged view for the (006) Bragg reflections. (c) Powder XRD patterns for some of (Li, Fe)OHFeSe single crystals.

Figure 2

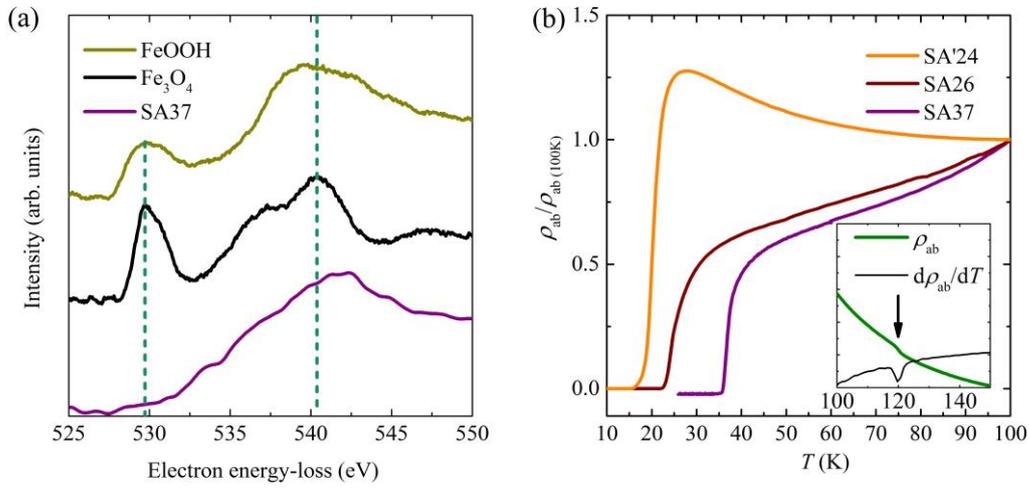

Fig. 2. (a) Oxygen K edges in the electron energy-loss spectroscopy (EELS) for (Li, Fe)OHFeSe (SA37), $Fe_3O_4$ and FeOOH, respectively. In the EELS measurement, the semi-convergence angle of the electron beam and the spectrum collection angle were estimated to be 10 and 30 mrad, respectively. (b) Temperature dependence of reduced in-plane electrical resistivity near the superconducting transition, for several representative samples. The inset shows a corresponding upward kink at $T_{afm}$, indicated by the arrow, in the in-plane electrical resistivity curve for a non-superconducting sample.

Figure 3

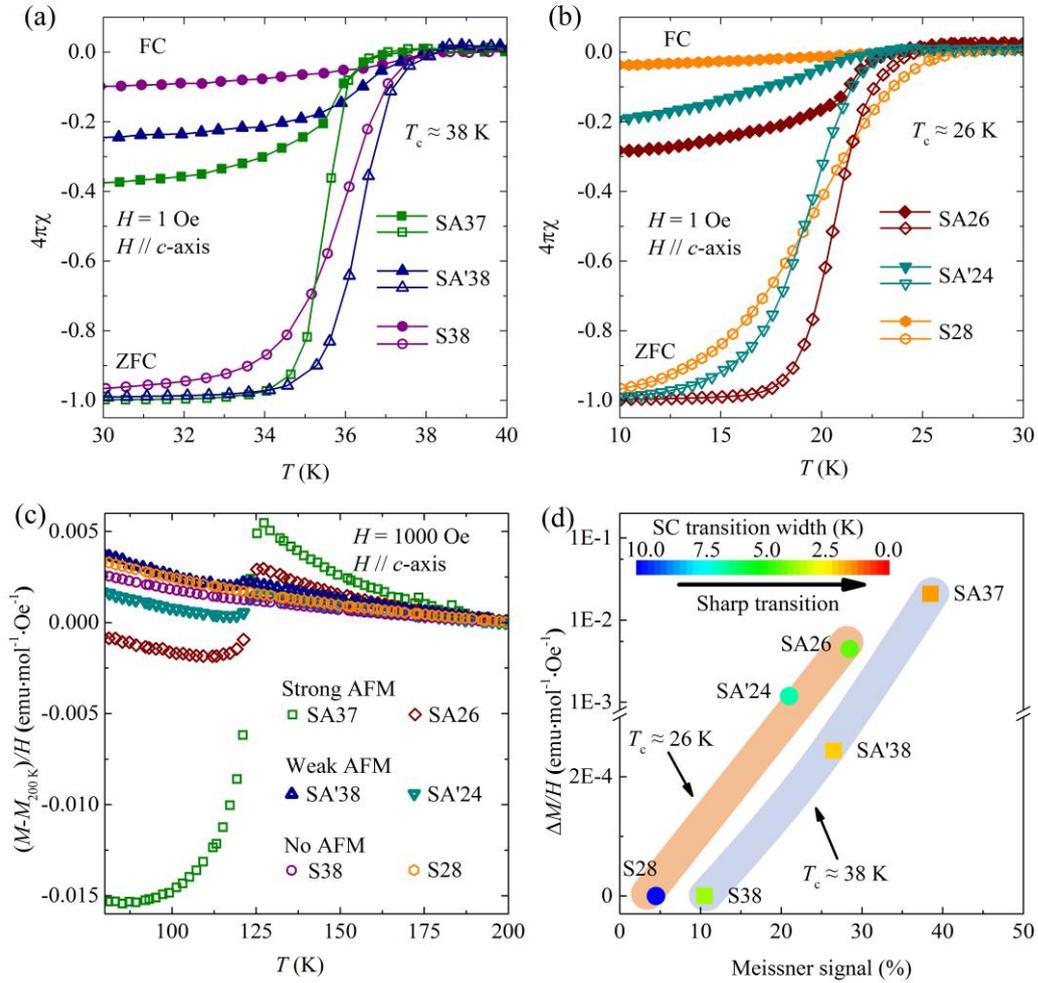

Fig. 3. Temperature dependence of static magnetic susceptibility near the
superconducting transitions, for the two sets of SA-/SA'-/S-samples with the
respective $T_c$'s of ~38 K (a) and ~26 K (b). The magnetic susceptibilities are corrected
for demagnetization factor. (c) Temperature dependence of reduced static
magnetization near $T_{afm}$ (~125 K), for the two sets of SA-/SA'-/S-samples with the
respective $T_c$'s ~38 K and ~26 K. $M_{200 K}$ represents the magnetization at 200 K. The
magnetization data for samples S41 and S40 are nearly the same as S38 and S28. And
the AFM signal at $T_{afm}$ for the non-SC sample (NSC) is stronger than the
superconducting samples. For clarity these data are not shown here. The
measurements were done in zero-field-cooling (ZFC) mode. (d) Plot of the AFM
signal size, i.e. $\Delta M/H$ near $T_{afm}$, versus the SC Meissner signal size, with the SC
transition width (between the 10% and 90% shielding signals) presented by the color
spectrum, for the two sets of SA-/SA'-/S-samples with the respective $T_c$'s ~38 K and
~26 K. The corresponding data are from (a), (b), and (c).

Figure 4

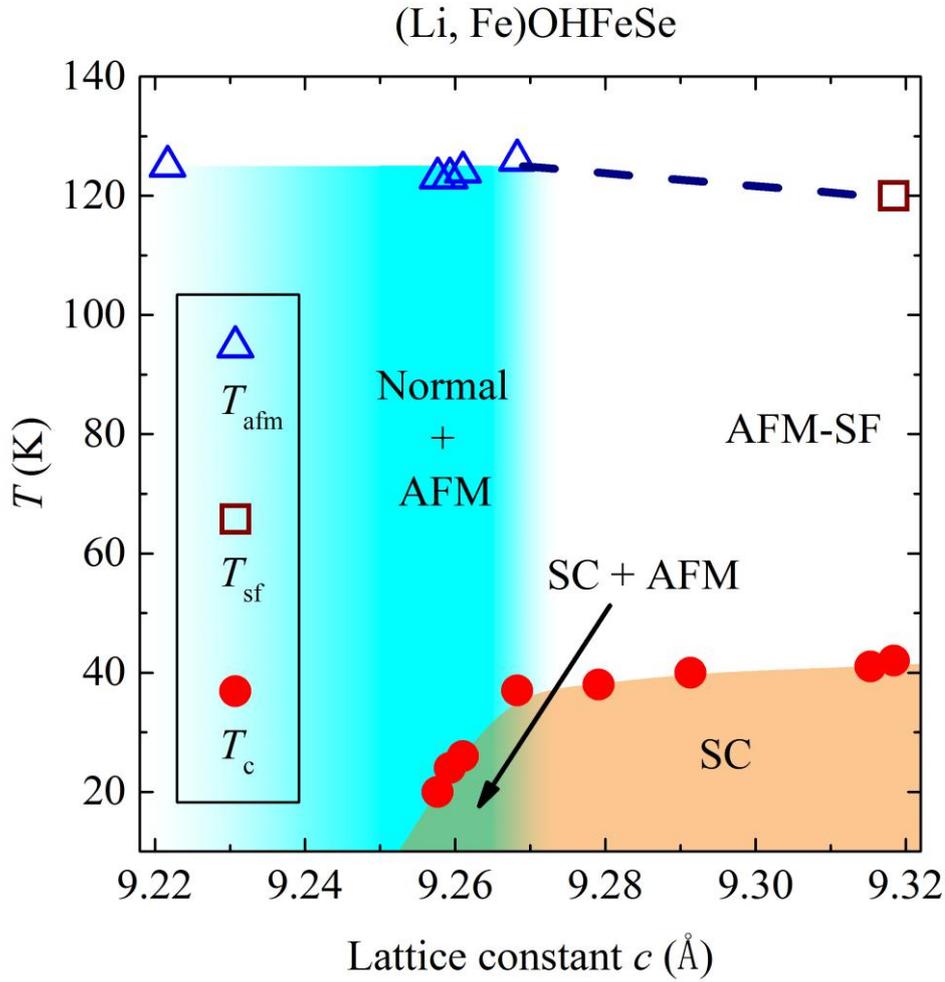

Fig. 4. Electronic phase diagram for (Li, Fe)OHFeSe superconductor system. The blue hollow triangles represent the coexisting three-dimensional AFM states (mesoscopic-to-macroscopic phase separations) below $T_{afm}$, the brown hollow square indicates the occurrence of two-dimensional AFM-SF below $T_{sf}$,[25] and the red solid circles are the $T_c$'s of the samples. In the left azure shaded area, the occurrence of a mesoscopic-to-macroscopic phase separation is sample and synthesis condition dependent. In the case of no AFM transition at $T_{afm}$, the phase separation is in a microscopic scale. The dashed blue line is an extrapolation from the right two-dimensional AFM-SF (below $T_{sf}$)[25] based on the present observations, overlapping with the microscopic phase separation region.

Table 1. $T_c$, $T_{afm}$, and unit cell parameters for some of the (Li, Fe)OHFeSe samples.

| Sample | $T_c$ (K) | $T_{afm}$ (K) | $a$ (Å) | $c$ (Å) |
| --- | --- | --- | --- | --- |
| S42[25] | 42 | / | 3.7827(4) | 9.3184(7) |
| S41 | 41 | / | 3.7811(1) | 9.3153(3) |
| S40 | 40 | / | 3.7816(2) | 9.2913(3) |
| S38 | 38 | / | 3.7872(2) | 9.2790(3) |
| SA37 | 37 | 126 | 3.7857(1) | 9.2682(3) |
| SA26 | 26 | 124 | 3.7887(1) | 9.2610(3) |
| SA'24 | 24 | 123 | 3.7942(2) | 9.2593(4) |
| SA20 | 20 | 123 | 3.78963(4) | 9.2577(1) |
| NSC | / | 125 | 3.7962(2) | 9.2217(2) |